\documentclass[amstex,aps,showpacs, twocolumn]{revtex4}%
\usepackage{graphicx}
\usepackage{color}
\usepackage{bm}
\usepackage{longtable}
\usepackage{amsmath}
\usepackage{amsfonts}
\usepackage{amssymb}%
\usepackage{lmodern}

\begin{document}
\title{Disembodiment of arbitrary number of properties in  quantum Cheshire cat experiment }
\author{A. K. Pan \footnote{akp@nitp.ac.in}}
\address{National Institute of Technology Patna, Ashok Rajpath, Patna 800005, India}
\begin{abstract}
The quantum Cheshire cat is an effect demonstrated within the framework of weak measurement aided with post-selection scenario where the property (say, grin) of a quantum particle (the cat) can be found in a spatially separated location from its position. In this work, we first propose interesting variants of quantum Cheshire cat where disembodiment of two different properties of a `cat' is demonstrated, so that, the \emph{grin} and the \emph{meowing} of it can be found in two different paths on an interferometer while the cat is in another path. We further extend this proposal for three-qutrit and $d$-qudit systems of a single particle. We provide sketches of the experimental proposals for testing our scheme by using the path, spin and energy degrees of freedoms of a single neutron and  the path, polarization and orbital angular momentum of a single photon. 
\end{abstract}
\maketitle
The idea of a weak measurement aided with post-selection (henceforth, WMPS) was introduced in 1988 by Aharonov et al. \cite{aav}. In contrast to the strong measurement, in WMPS scenario, the average value of an observable (historically termed as weak value) can produce results beyond the eigenvalue spectrum of a given observable.  Such concept of quantum measurement has allowed the development of novel and important experimental techniques aided with quantum mechanics. A flurry of theoretical [2-28] and experimental works [29-48] on WMPS have been reported in last two decades. The eccentric weak value induced  effect is shown to be useful in many practical applications, such as, identifying tiny spin Hall effect \cite{hosten}, detecting very small transverse beam deflections \cite{starling}, measuring average quantum trajectories for photons \cite{steinberg11}, improving signal-to-noise ratio for determination of small phase through interferometry \cite{str}, protecting a quantum state \cite{kim} and many more. Before proceeding further, let us first recapitulate the essence of WMPS scenario.

Consider a system is prepared in a state (commonly known as pre-selected state)   $|\chi_{i}\rangle$ and $\hat{O}$ is the observable to be measured on the system. If the measurement interaction between the system and the apparatus is weak, the system state remains grossly undisturbed. Now, after the weak measurement interaction if a particular outcome  $|\chi_{f}\rangle$ is selected out by using a strong measurement sequentially, the final pointer state yields the weak value, quantified by the formula \cite{aav} is given by
\begin{equation}
(O)_{w}=\frac{\left\langle \chi_{f}|O|\chi
_{in}\right\rangle }{\left\langle \chi_{f}\right\vert \chi_{in}\rangle
}\label{aavwvalue},%
\end{equation}
The weak value $(O)_{w}$ can be outside the range of eigenvalues and can even be complex. For example, if $|\chi_{in}\rangle= \alpha |\uparrow_{z}\rangle+ \beta|\downarrow_{z}\rangle$ (where $\alpha$ and $\beta$ in general complex satisfying $|\alpha|^2+|\beta|^2=1$), the post-selected state is $|\chi_{f}\rangle=\left\vert \uparrow
\right\rangle _{z}$ and the observable $\hat{O}= \hat{\sigma}_{x}$, the the weak value is given by $(\sigma_{x})_{w}=\beta/\alpha$ which is in general complex. The value of $(\sigma_{x})_{w}$ can become arbitrarily large while $\alpha$ approaches to zero \cite{aav}. The physical interpretation and implications of complex and large weak values have been widely discussed in the literature (see, for example, \cite{dressel}). 

Beside novel practical applications, the WMPS also provides new insights into conceptual foundations of quantum mechanics \cite{av91,vaidman,aha02,pussey,mit,vaid13,danan,aha17}. One of such striking examples is a work by Arahonov, Popescu and Skrzypczyk \cite{ah13}, who demonstrated the quantum version of the mystic Cheshire cat. Such a cat was first appeared in Lewis Carroll's novel, whose grin was disembodied from its body. Using the notion of WMPS, the quantum version of Cheshire cat is demonstrated \cite{ah13} in a suitable Mach-Zehender interferometric setup where the polarization of a single photon can be found in a path, different from the path taken by photon. Further studies have been made along this directions \cite{french,gur,mat,du17}. A general scheme for testing three-box paradox and Cheshire cat paradox is  presented \cite{mat} for the case of spin-1 atoms. It is also shown that the grin can also be exchanged if disembodiment of two photons from their respective polarizations is considered in a suitable setup \cite{pati}. The proposal of \cite{ah13} has recently been experimentally tested  by using single neutron \cite{denk} and single photon \cite{ashby,kim1}. However, a string of criticisms have been made on the very concept of quantum Cheshire cat and the experiment\cite{corr15}. 

Let us now  encapsulate the \emph{gadanken} experiment demonstrated in \cite{ah13} demonstrates the quantum Chesire cat effect.  For this, consider a single photon is pre-selected in a state at time $t_{i}$ is given by 
\begin{align}
	|\Psi_{i}\rangle=\frac{1}{\sqrt{2}}\left(|\psi_{1}\rangle |1\rangle+ |\psi_{2}\rangle |0\rangle\right)
\end{align}
where $|\psi_1\rangle$ and $|\psi_2\rangle$ can be considered as orthogonal path states of the Mach-Zehender interferometer and the states  $|0\rangle\equiv (1 \ 0)^T$ and $|1\rangle\equiv (0 \ 1)^T$ representing the polarization of the photon, are the eigenstates of the dichotomic Pauli observable $\hat{\sigma}_{z}$.

The photon is post-selected in later time $t_{f}$ ($t_f > t_i$) in a state is given by  
\begin{align}
	|\Psi_{f}\rangle=\frac{1}{\sqrt{2}}\left(|\psi_{1}\rangle + |\psi_{2}\rangle \right)|1\rangle
\end{align}
Given the pre- and post-selection scenario, to inspect in which of the two paths the photon passes, the weak measurements of the path observables $\Pi_{1}=|\psi_{1}\rangle\langle \psi_{1}|$ and $\Pi_{2}=|\psi_{2}\rangle\langle \psi_{2}|$ are considered at $t$ with  $t_f > t > t_i$. The weak value of the observables $\Pi_{1}$ and $\Pi_{2}$ are given by
\begin{align}
	(\Pi_{1})_w =1; (\Pi_{2})_w =0
\end{align}
which imply that for a successful post-selection the particle passes through the path $|\psi_1\rangle$. In order to inspect in which path $|\psi_{i}\rangle$ (with $i=1,2$) the polarization can be found, one chooses the joint observables  $\Pi_{i}\sigma_{x}^{p}$ where the polarization observable is a Pauli operator $\sigma_{x}^{p}=|+\rangle\langle +|-|-\rangle\langle -|$ with $|\pm\rangle=(|0\rangle+|1\rangle)/\sqrt{2}$. The weak values of those joint path-polarization observables are 
\begin{align}
	(\Pi_{1}\sigma_{x}^{p})_w =0; (\Pi_{2}\sigma_{x}^{p})_w =1   
\end{align}
 Surprisingly, the polarization of the photon travels through the path $|\psi_{2}\rangle$ although the photon took the other path, as if the polarization property is disembodied from the photon. 

In this paper, we provide proposals to demonstrate the disembodiment of two or more properties of a single particle from its matter.  We have further generalized our proposal for $d$-qudit systems of a single particle for disembodiment of $d-1$ number of properties from the particle. For the disembodiment of two properties, we provide sketches of possible experimental proposals that can be implemented by using the path, spin and energy degrees of freedom of a single neutron, and the path, polarization and orbital angular momentum degrees of freedom of a single photon.

\section{Disembodiment of multiple properties from matter}

By extending the above idea, here we first demonstrate the disembodiment of two different properties of a single particle. We further generalize our proposal for demonstrating the spatial separations of arbitrary number of properties of a single particle where each of the properties corresponds to a dichotomic observable. For our purpose, we specifically choose the pre-selected state of the particles as 
\begin{align}
	|\Psi_{i}\rangle=\frac{1}{\sqrt{3}}\left(|\psi_{1}\rangle |11\rangle+ |\psi_{2}\rangle |01\rangle+|\psi_{3}\rangle |10\rangle\right)
\end{align}
where $|\psi_{i}\rangle$ ($i=1,2,3$) are the path states of a three-path interferrometer and $|11\rangle \equiv |1\rangle^{1}\otimes|1\rangle^{2}$ represents the state of the property-1 and property-2 of the same single particle. 
The post-selected state is taken to be
\begin{align}
	|\Psi_{f}\rangle=\frac{1}{\sqrt{3}}\left(|\psi_{1}\rangle + |\psi_{2}\rangle+|\psi_{3}\rangle \right)|11\rangle
\end{align}
Note that, many such choices of pre- and post-selection are possible that can provide the similar disembodiment of properties. Now, for inspecting the path taken by the particle in this a pre- and post-selection scenario we consider the weak value of the path projectors, $\Pi_{1}=|\psi_{1}\rangle\langle \psi_{1}|$, $\Pi_{2}=|\psi_{2}\rangle\langle \psi_{2}|$ and $\Pi_{3}=|\psi_{3}\rangle\langle \psi_{3}|$ are the following
\begin{align}
	(\Pi_{1})_w =1; (\Pi_{2})_w =0; (\Pi_{3})_w =0  
\end{align}
This then implies that the particle passes through the path $|\psi_{1}\rangle$. In order to examine the path in which the property-1  travels, we consider the weak value of the following joint observables  $\Pi_{1}\sigma_{x}^{1}$, $\Pi_{2}\sigma_{x}^{1}$ and $\Pi_{3}\sigma_{x}^{1}$ providing 
\begin{align}
	(\Pi_{1}\sigma_{x}^{1})_w =0; (\Pi_{2}\sigma_{x}^{1})_w =1; (\Pi_{3}\sigma_{x}^{1})_w =0   
\end{align}
meaning that the  property-1 is found in path $|\psi_{2}\rangle$. If we calculate the weak values of 
 $\Pi_{1}\sigma_{x}^{2}$, $\Pi_{2}\sigma_{x}^{2}$ and $\Pi_{3}\sigma_{x}^{2}$ we have
\begin{align}
	(\Pi_{1}\sigma_{x}^{2})_w =0; (\Pi_{2}\sigma_{x}^{2})_w =0; (\Pi_{3}\sigma_{x}^{2})_w =1   
\end{align}
reveals that the property-2 travels through the path $|\psi_{3}\rangle$.

Interestingly, if we jointly inspect the property-1 and property-2 by weakly measuring the observables $\Pi_{{i}}\sigma_{x}^{1} \sigma_{x}^{2}$ where $i=1,2,3$ we obtain
 \begin{align}
	(\Pi_{1}\sigma_{x}^{1}\sigma_{x}^{2})_{w}=(\Pi_{2}\sigma_{x}^{1}\sigma_{x}^{2})_{w}=(\Pi_{3}\sigma_{x}^{1}\sigma_{x}^{2})_{w}=0 
\end{align}
i.e., both the properties cannot be found jointly in any path although they are found individually in different paths.
 
The above result can straightforwardly be extended for disembodying $n-1$ number of properties where the path degree of freedom should be represented by $n-$level system and properties can still be taken as dichotomic system. For example, if one takes the pre-selected state
\begin{eqnarray}
	|\Psi_{i}\rangle &=& 	\frac{1}{\sqrt{N}}(|\psi_{1}\rangle |1..1\rangle+ |\psi_{2}\rangle |01..1\rangle\\
	\nonumber
	&+&|\psi_{3}\rangle |101..1\rangle..+|\psi_{n}\rangle |1..10\rangle)
\end{eqnarray}
the post-selected state
 \begin{align}
	|\Psi_{f}\rangle=\frac{1}{\sqrt{N}}\left(|\psi_{1}\rangle + |\psi_{2}\rangle+|\psi_{3}\rangle +..+|\psi_{n}\rangle\right)|1..1\rangle
\end{align}
and $n$-number suitable weak measurements of the relevant observables along $n$-paths, the disembodiment of the $n-1$ properties can be shown straightforwardly. However, the experimental testing of the disembodiment of more than two  properties could be a huge challenge. 

Note that, upto now we have considered an $n$-path interferometer for the disembodiment of $n-1$ number of properties. Also, we assumed that the properties can be represented as a dichotomic observables. Next we demonstrate a scheme to show the disembodiment of $d-1$ number of properties where the properties are also represented as a $d$-outcome observables. We first demonstrate the scheme for disembodying two properties represented by three-outcome observables by using a three path interferometer. This is further generalized for arbitrary $d$ number of qudits.
\section{Proposals for  $3$-qutrits and its extension to $d$-qudits}    
As already indicated, we use a three-path interferometer which is mathematically a qutrit system and two properties  also correspond to qutrit observable.  We consider the pre-selected state of the particle is of the form
\begin{equation}
|\Psi_{i}\rangle= \frac{1}{\sqrt{3}}\left(|\psi_1\rangle|11\rangle+|\psi_2\rangle|22\rangle+|\psi_3\rangle|33\rangle\right),
\end{equation}
where $|1\rangle=\left(1 0 0\right)^{T}$, $|2\rangle=\left(0 1 0\right)^{T}$ and $|3\rangle=\left(0 0 1\right)^{T}$ with $|33\rangle\equiv|3\rangle^{1}\otimes |3\rangle^{2}$ corresponds to property-1 ans 2. Let the particle is post-selected in the state given by 
\begin{equation}
|\Psi_{f}\rangle=\frac{1}{\sqrt{3}}(|\psi_1\rangle|11\rangle+|\psi_2\rangle|12\rangle+|\psi_3\rangle|31\rangle).
\end{equation}

Given the pre- and post-selection in Eqs. (14) and (15), we first examine in which path the particle passes by using the weak measurements of the  projectors  corresponding to the  three paths. This yields 
\begin {equation}
(\Pi_{1})_{w}=1,(\Pi_{2})_{w}=(\Pi_{3})_{w}=0,
\end{equation}
i.e.,  the particle can only be found in path $|\psi_1\rangle$ for the above pre- and post-selected experiment. Now, we turn our attention to the measurements of the property-1 and property-2 separately along the three paths. A suitable observable $\Pi_{i}{\hat J}^{(1)}$ with $i=1,2,3$ is chosen to probe the localization of the property-1, where
\begin{equation}
{\hat J}^{(1)}= \left(|1\rangle\langle 2|+|2\rangle\langle 1|\right)
\end{equation} 
is a qutrit observable. We then have the following results, 
\begin{equation}
\left(\Pi_{1}{\hat J}^{(1)}\right)_{w}=0; \left(\Pi_{2}{\hat J}^{(1)}\right)_{w}=1; \left(\Pi_{3}{\hat J}^{(1)}\right)_{w}=0
\end{equation}
showing the property-1  is found in the path $|\psi_2\rangle$, although the particle was in the path $|\psi_1\rangle$. Similarly, one can examine the path taken by the  property-2 by weakly measuring  $\Pi_{i}{\hat J}^{(2)}$) with $i=1,2,3$, where
\begin{equation}
\nonumber
{\hat J}^{(2)}=  \left(|1\rangle\langle 3|+|3\rangle\langle 1|\right)
\end{equation}
 one finds
\begin{equation}
\left(\Pi_{1}{\hat J}^{(2)} \right)_{w}=\left(\Pi_{2}{\hat J}^{(2)} \right)_{w}=0; \left(\Pi_{3}{\hat J}^{(2)} \right)_{w}=1.
\end{equation}
Hence, the property-2 is found in the channel $|\psi_3\rangle$, but the particle and the property-1 were found in path $|\psi_1\rangle$ and $|\psi_2\rangle$ respectively.

The above results derived for three three-qutrit systems can be generalized $d$ number of qudit systems. In such a case, the pre-selected state can be taken as\begin{equation}
|\psi_{i}\rangle= \frac{1}{\sqrt{d}}\left(|\psi_1\rangle|1....1\rangle+|\psi_2\rangle|2....2\rangle+....+|\psi_d\rangle|d....d\rangle\right)
\end{equation}
where $|1\rangle=\left(1 0 . . . . 0\right)^{T}$, $|2\rangle=\left(0 1 0 . . . . 0\right)^{T}$ and $|d\rangle=\left(0 . . . . 0 1\right)^{T}$. 
The particle is post-selected in the state given by 
\begin{equation}
|\psi_{f}\rangle=\frac{1}{\sqrt{d}}(|\psi_1\rangle|11..1\rangle+|\psi_2\rangle|12...2\rangle+.....+|\psi_d\rangle| d...d 1\rangle)
\end{equation}

Using the pre- and post-selections given by Eqs. (20) and (21), the weak measurements of the path projectors $\{\Pi_{k}\}$ with $k=1,2,3...d$  provide 
\begin {equation}
(\Pi_{k=1})_{w}=1; (\Pi_{k\neq 1})_{w}=0
\end{equation}
Then the particle can only be found in path $|\psi_1\rangle$ for the above pre- and post-selected scenario. To find the $1^{st}$ property, we measure the same observable given by Eq.(17), so that
\begin{eqnarray}
 (\Pi_{k=2}{\hat J}^{(1)})_{w}=1; (\Pi_{k\neq2}{\hat J}^{(1)})_{w}=0
\end{eqnarray}

Similarly, for the observable $\Pi_{d}{\hat J}^{(d-1)}$, we have
\begin{equation}
(\Pi_{k=d}{\hat J}^{(d-1)})_{w}=1; (\Pi_{k\neq d}{\hat J}^{(d-1)})_{w}=0
\end{equation}
where 
\begin{equation}
{\hat J}^{(d-1)}= \left(|1\rangle\langle d|+|d\rangle\langle 1|\right)
\end{equation}

Hence, the scheme described here demonstrates the disembodiment of $d-1$ number of properties of a single particle.

\section{Proposed experimental proposals using photon and neutron}
We provide sketches of the two experimental proposals for testing the schemes for disembodying two different properties from the quantum matter. The first proposal is by using single photon where disembodiment of polarization and orbital angular momentum degrees of freedom from the photon can be achieved. In the second proposal, we propose the disembodiment of spin and energy degrees of freedom from a single neutron. The  requirements for experimentally implementing the schemes are already available for single photon in \cite{nagali} and for single neutron in \cite{yuji}.

The scheme utilizing the single photon can be implemented by using beam splitters (BSs), mirrors (M1 and M2), half wave plates (HWPs), phase shifters (PSs), Q-plates (QPs) and three detectors (D1, D2 and D3) as depicted in Fig.1.  We use two symmetric (1/2:1/2) beam splitters (BS2 and BS3) and two asymmetric (1/3:2/3) beam splitters (BS1 and BS4). HWP converts left circularly polarize state ($|L\rangle$) to right circularly polarize state ($|R\rangle$) and the PS introduces a phase shift of amount $\pi$. A QP acts on both polarization and orbital angular momentum states (say,$|m=0\rangle$), so that, $|R\rangle|m=0\rangle\rightarrow |L\rangle|m=-2\rangle$ and $|L\rangle|m=0\rangle\rightarrow |R\rangle|m=+2\rangle$. The procedure to prepare the state given by Eq.(6) is the following.  The photon passes through a asymmetric$(1/3:2/3)$ beam splitter (BS1) so that the state after the BS1 can be written as 
\begin{align}
	|\Psi_{BS1}\rangle=\frac{1}{\sqrt{3}}\left(|\psi_{1}\rangle + i\sqrt{2}|\psi^{\prime}_{1}\rangle \right)|L\rangle|0\rangle
\end{align}
where $|0\rangle\equiv|m=0\rangle$ denotes the orbital angular momentum state. A symmetric (1/2:1/2) beam splitter (BS2) along the path $|\psi^{\prime}_{1}\rangle $  splits it to $|\psi_{2}\rangle $ and $|\psi_{3}\rangle $. By using the two phase shifters along the paths $|\psi_{2}\rangle $ and $|\psi_{3}\rangle ,$ and by taking into the $\pi/2$ phase shift produced by the mirror (M1), we obtain
\begin{align}
	|\Psi_{BS2}\rangle=\frac{1}{\sqrt{3}}\left(|\psi_{1}\rangle + |\psi_{2}\rangle+|\psi_{3}\rangle \right)|L\rangle|0\rangle
\end{align}
\begin{figure}[h]
{\rotatebox{0}{\resizebox{8.5cm}{6.0cm}{\includegraphics{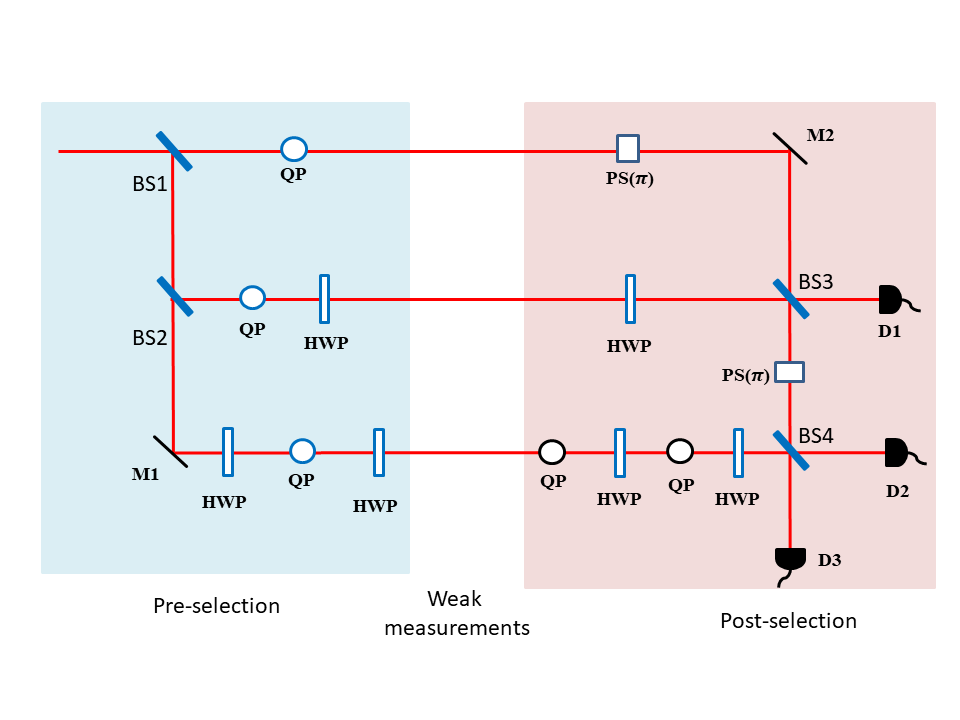}}}}
\caption{\footnotesize The photon is prepared in a state $|\Psi_i\rangle$ by using beam splitters, half wave plates to manipulate the polarization and Q-plates to manipulate both the polarization and orbital angular momentum. The post-selection process is arrange in a manner that for a successful post-selection of the system in a state  $|\Psi_f\rangle$ the detectors D1 and D2 never click and only D3 will click.}
\end{figure}
Next, the following operations, such as, Q-plate (QP) along the path $|\psi_{1}\rangle$, both QP and HWP along the path $|\psi_{2}\rangle$ and HWP, QP and HWP along the path $|\psi_{3}\rangle$ provide the state is given by
 \begin{align}
	|\Psi_{i}\rangle=\frac{1}{\sqrt{3}}\left(|\psi_{1}\rangle|R\rangle|+2\rangle + |\psi_{2}\rangle|L\rangle|+2\rangle+|\psi_{3}\rangle |R\rangle|-2\rangle\right)
\end{align}
which is our pre-selected state and equivalent to Eq. (6).
 
For post-selecting the photon in a state as given by Eq. (7), a suitable arrangement is required that comprises a PS, three HWPs, two QPs, a symmetric BS3 (1/2:1/2), a asymmetric BS4 (2/3:1/3), a mirror (M2) and three detectors (D1, D2 and D3). A joint action of PS (defined in the basis of $|\psi_{1}\rangle, |\psi_{2}\rangle, |\psi_{3}\rangle$  ) and $M_2$ in the state $|\psi_{1}\rangle|R\rangle|+2\rangle$ provides $-i|\psi_{1}\rangle|R\rangle|+2\rangle$  and HWP in $|\psi_{2}\rangle|L\rangle|+2\rangle $ produces $|\psi_{2}\rangle|R\rangle|+2\rangle $. Now, the joint actions of QPs and HWPs in $|\psi_{3}\rangle |R\rangle|-2\rangle$ gives $|\psi_{3}\rangle |R\rangle|2\rangle$. Then the state just before the BS3 and BS4 can be written as

\begin{align}
	|\Psi^{\prime}\rangle=\frac{1}{\sqrt{3}}\left(-i|\psi_{1}\rangle+ |\psi_{2}\rangle+|\psi_{3}\right) |R\rangle|+2\rangle
\end{align}

After passing through the symmetric BS3 $(1/2:1/2)$ and asymmetric BS4 $(1/3:2/3)$, the state is given by

\begin{align}
	|\Psi^{\prime\prime}\rangle=\frac{1}{\sqrt{3}}\left(|D_{1}\rangle+ |D_{2}\rangle+|D_{3}\rangle\right) |R\rangle|+2\rangle
\end{align}
  where 
	\begin{align}
		|D_{1}\rangle=\frac{1}{\sqrt{2}} \left(-|\psi_{1}\rangle+ |\psi_{2}\rangle\right)|R\rangle|+2\rangle
	\end{align}
\begin{align}
		|D_{2}\rangle= \frac{1}{\sqrt{6}} \left(i|\psi_{1}\rangle+ i|\psi_{2}\rangle +2|\psi_{3}\rangle\right)|R\rangle|+2\rangle
\end{align}
 and 
	
	\begin{align}
|\Psi_{f}\rangle\equiv |D_{3}\rangle= \frac{1}{\sqrt{3}}\left(|\psi_{1}\rangle+ |\psi_{2}\rangle +|\psi_{3}\rangle\right)|R\rangle|+2\rangle
	\end{align}
which is the desired post-selected state. Note that a global phase of $\pi/2$ is omitted in $|D_{3}\rangle$.  For our purpose of disembodiment of properties from the particle, we consider the detection at the D3 detector only, i.e., in the state  $|\Psi_{f}\rangle$.

To obtain the disembodiment of polarization and orbital angular momentum, one can consider weakly measuring $\Pi{_i}$, $\Pi_{i}\sigma_{x}^{p}$ and $\Pi_{i}\sigma_{x}^{l}$ with $i=1,2,3$ and $\sigma_{x}^{p}$ and $\sigma_{x}^{l}$ observables corresponding to the polarization and orbital angular momentum as provided in Eqs. (8-10).

A similar experimental arrangement as in Fig. 1 can be made by using neutron interferometric setup in \cite{yuji}. In this case, instead of QP the radio-frequency (RF) oscillator with variable frequencies ($\omega$) can be used to jointly manipulate the spin and the energy degrees of freedom, so that, $RF^{\pi}_{\omega}$ converts $\rangle |\uparrow\rangle|E_{0}\rangle$ to $i|\downarrow\rangle|E_{0}-\hbar\omega\rangle$ where $|\uparrow\rangle $ and $|E_{0}\rangle$  are the spin and energy states of the neutron respectively.  Let the initial state entering the BS1 is $|\psi_{0}\rangle|\uparrow\rangle|E_{0}\rangle$ where $|\psi_{0}\rangle$, $|\uparrow\rangle$ and $|E_{0}\rangle$ are the path, spin and total energy state of the neutron respectively. Now, the state after the asymmetric BS1 is given by   
\begin{align}
|\Psi_{BS1}\rangle=\frac{1}{\sqrt{3}}\left(|\psi_{1}\rangle + i\sqrt{2}|\psi^{\prime}_{1}\rangle \right)|\uparrow\rangle|E_{0}\rangle
\end{align}   
The state $|\psi^{\prime}_{1}\rangle$ is subject to the   BS2 (1/2:1/2) which spits the state into $|\psi_{2}\rangle$ and $|\psi_{3}\rangle$. By successively placing a PS and a spin-flipper along the channel $|\psi_{2}\rangle$, and by placing a  oscillator $RF^{\pi}_{\omega}$ followed by a another spin-flipper along the path $|\psi_{3}\rangle$, we obtain the state, is given by
\begin{align}
|\Psi_{i}\rangle=\frac{1}{\sqrt{3}}\left(|\psi_{1}\rangle |\uparrow\rangle|E_{0}\rangle+|\psi_{2}\rangle |\downarrow\rangle|E_{0}\rangle+|\psi_{3}\rangle |\uparrow\rangle|E_{0}-\hbar\omega\rangle\right)
\end{align}   

The post-selection can be done using the similar procedure adopted for the case of photon is given by
\begin{align}
|\Psi_{f}\rangle=\frac{1}{\sqrt{3}}\left(|\psi_{1}\rangle +|\psi_{2}\rangle +|\psi_{3}\rangle \right)|\uparrow\rangle|E_{0}\rangle
\end{align}   

One can now consider weakly measuring $\Pi_{i}$, $\Pi_{i}\sigma_{x}^{s}$ and $\Pi_{i}\sigma_{x}^{e}$ with $i=1,2,3$ and $\sigma_{x}^{s}$ and $\sigma_{x}^{e}$ observables corresponding to the spin and energy degrees of freedom and can obtain the disembodiment of spin and energy degrees of freedom from the single neutron as given by Eqs.(8-10).

Note that, above schemes are not specific to the asymmetric beam splitters. One may use 50:50 BS but for this case the suitable choices of pre- and post-selected states need to be made. For testing three-qutrits scheme presented in Sec. III, a spin-1 atom can also be used in which spin angular momentum and energy degrees of freedom can be disembodied. 
\section{Summary and Conclusions}
In summary, we proposed variants of quantum Cheshire cat where two or more properties of a single particle can be disembodied. The proposal is further extended for disembodiment of $d-1$ number of properties by using $d$-qudits systems of a single particle. However, experimental generation of a $d$-qudit systems of a single particle with $d>3$ requires adequate advanced technology which may not be available currently. We provided the sketches of the possible experimental schemes for testing the disembodiment of two properties from the particle by using  path, polarization and orbital angular momentum degrees of freedom of a single photon, and the path, spin and energy degrees of freedom of a single neutron.  Specifically, we propose that how the polarization and orbital angular momentum properties can be spatially separated from the photon, and the spin and energy degrees of freedom can be separated from a single neutron. The experimental proposals sketched here can be tested using the existing technologies as already developed in \cite{nagali,yuji}.

{\bf Acknowledgments:} Author acknowledges the support from the project DST/ICPS/QuEST/2018/Q-42.

\end{document}